\documentclass[12pt,draftcls, onecolumn]{IEEEtran}

\usepackage{amsmath}
\usepackage{amssymb}
\usepackage[dvips]{graphicx}
\usepackage{setspace}
\usepackage{epsfig}
\usepackage{amsmath}
\usepackage{amssymb}
\usepackage[dvips]{graphicx}
\usepackage{epsfig}

\newcommand{\ffigsize}{0.6}
\newcommand{\figsize}{1}
\newcommand{\tSNR}{\text{SNR}}
\newcommand{\E}{\mathbb{E}}

\newtheorem{theo}{Theorem}

\begin{document}
\title{Effective Capacity Analysis of Cognitive Radio Channels for Quality of Service Provisioning}
\author{\authorblockN{Sami Akin and
Mustafa Cenk Gursoy}
\thanks{The authors are with the Department of Electrical
Engineering, University of Nebraska-Lincoln, Lincoln, NE, 68588
(e-mails: samiakin@huskers.unl.edu, gursoy@engr.unl.edu).}
\thanks{This work was supported by the National Science Foundation under Grants CCF -- 0546384 (CAREER) and CNS -- 0834753.}}

\date{}

\maketitle

\vspace{-.7cm}
\begin{abstract}

In this paper, cognitive transmission under quality of service (QoS) constraints is studied. In the cognitive radio channel model, it is assumed that the secondary transmitter sends the data at two different average power levels, depending on the activity of the primary users, which is determined by channel sensing performed by the secondary users. A state-transition model is constructed for this cognitive transmission channel. Statistical limitations on the buffer lengths are imposed to take into account the QoS constraints. The maximum throughput under these statistical QoS constraints is identified by finding the effective capacity of the cognitive radio channel. This analysis is conducted for fixed-power/fixed-rate, fixed-power/variable-rate, and variable-power/variable-rate transmission schemes under different assumptions on the availability of channel side information (CSI) at the transmitter. The impact upon the effective capacity of several system parameters, including channel sensing duration, detection threshold, detection and false alarm probabilities, QoS parameters, and transmission rates, is investigated. The performances of fixed-rate and variable-rate transmission methods are compared in the presence of QoS limitations. It is shown that variable schemes outperform fixed-rate transmission techniques if the detection probabilities are high. Performance gains through adapting the power and rate are quantified and it is shown that these gains diminish as the QoS limitations become more stringent.

\end{abstract}

\begin{spacing}{1.75}

\section{Introduction}
With the rapid growth in the wireless networks in the last two decades, the scarcity in
spectrum has become a serious problem for spectrum sharing, since much of the prime wireless
spectrum has been allocated for specific applications. However, recent measurements show
that the licensed spectrum is severely under-utilized. This has caused a tremendous interest in
using the spectrum dynamically by exploring the empty spaces in the spectrum without disturbing
the primary users. In such systems, in order to avoid the interference to the primary users, it
is very important for the cognitive secondary users to detect the activity of the primary users.
When the primary users are active, the secondary user should either avoid using the channel or
transmit at low power in order not to exceed the noise power threshold of the primary users, whereas the secondary users can use the channel without any constraints when the
channel is free of the primary users.

With the above-mentioned motivation, recent years have witnessed a large body of work on channel sensing and dynamic spectrum sharing. Dynamically sharing the spectrum in the time-domain by exploiting whitespace between the bursty
transmissions of a set of users, represented by an 802.11b based wireless LAN (WLAN), is considered
by the authors in \cite{stefan}, where a model that describes the busy and idle periods of a WLAN is
considered. 
The authors in \cite{poor} investigated the problem of maximally utilizing the spectrum opportunities in cognitive radio networks
with multiple potential channels, and studied the optimal sensing order problem in multi-channel cognitive
medium access control with opportunistic transmission. In their paper \cite{yunxia}, Chen \textit{et al.}
developed an optimal strategy for opportunistic spectrum access (OSA) by integrating the design of spectrum
sensor at the physical layer with that of spectrum sensing and access policies at the medium access control (MAC) layer,
considering the maximization of the throughput of secondary users as the design objective while limiting their collisions with
primary users. The authors in \cite{qzhao} analyzed the problem of opportunistic access to parallel channels occupied
by the primary users under a continuous-time Markov chain modeling of the channel occupancy by the primary users, where they
proposed a slotted transmission strategy for secondary users using a periodic sensing strategy with optimal dynamic access.
Their objective was also to maximize the channel utilization of the secondary users while controlling their interference to
the primary users. Poor \textit{et al.} introduced a novel wideband spectrum sensing technique, called as multiband
joint detection in \cite{poor2}, that jointly detects the signal energy levels over multiple frequency bands rather than
considering one band at a time, which is proposed to be efficient in improving the dynamic spectrum utilization and
reducing interference to the primary users. In \cite{jindal}, the capacity of opportunistic
secondary communication over a spectral pool of two independent channels is explored and it is shown
that the benefits of spectral pooling are lost in dynamic spectral environments.

Note that spectrum sensing, which is crucial in the detection of the presence of primary users and hence in interference management, also induces a cost in terms of reduced time for data transmission. Motivated by this fact, the authors in \cite{liang} studied the tradeoff between channel sensing and throughput considering the Shannon capacity as the throughput metric. They formulated an optimization problem and identified the optimal sensing time which yields the highest throughput while providing sufficient protection in terms of interference to the primary users.

As described above, issues regarding channel sensing, spectrum sharing and throughput in cognitive radio networks have been extensively studied recently (see also for instance \cite{specialissue}). However, another critical concern of providing quality of service (QoS) guarantees over cognitive radio channels has not been sufficiently addressed yet.
In many wireless communication systems, providing certain QoS assurances is
crucial in order to provide acceptable performance and quality. However, this is a challenging task
in wireless systems due to random variations experienced in channel conditions and random fluctuations
in received power levels and supported data rates. Hence, in wireless systems, generally statistical,
rather than deterministic, QoS guarantees can be provided. Note that the situation is further exacerbated
in cognitive radio channels in which the access to the channel can be intermittent or transmission occurs at lower power levels depending on the activity of the primary users. Furthermore, cognitive radio
can suffer from errors in channel sensing in the form of false alarms. Hence, it is of paramount interest to analyze
the performance of cognitive radio systems under QoS limitations in the form of delay or buffer constraints.

The maximum throughput levels achieved in wireless systems operating under such statistical QoS constraints
can be identified through the notion of effective capacity. The effective capacity is defined in \cite{Wu} as the maximum
constant arrival rate that a given time-varying service process can support while meeting the QoS requirements. The application and analysis of effective capacity in various settings has attracted
much interest. When both the transmitter and the receiver have the information of the instantaneous channel gains,
Tang and Zhang in \cite{zhang} analyzed the effective capacity and derived the optimal power and rate adaptation
techniques which maximize the system throughput under QoS constraints. The effective capacity and efficient resource allocation
strategies for Markov wireless channel models are analyzed by Liu \textit{et al.} in \cite{LLui} where fixed-rate
transmission schemes are considered. In this study, the continuous-time Gilbert-Elliot channel with ON and OFF states
is studied. In \cite{gursoy}, the energy efficiency under QoS constraints is investigated
by analyzing the normalized effective capacity in the low-power and wideband regimes. In this work,
variable-power/variable-rate and fixed-power/variable-rate transmission schemes are considered assuming the availability of
channel side information at both the transmitter and receiver or only at the receiver.

In this paper, we study the effective capacity of cognitive radio channels in order to identify the performance in the presence of statistical QoS constraints. The secondary users are assumed to initially perform channel sensing to detect the
activity of the primary users and then transmit the data at two different average power levels depending on the presence or absence of active primary users. More specifically, the contributions of this paper are the following:
\begin{enumerate}
\item We identify a state-transition model for cognitive transmission by comparing the transmission rates with the instantaneous channel capacity values, and incorporating the sensing decision and its correctness into the model.

\item We determine the effective capacity of cognitive transmission and provide a tool for the performance analysis in the presence of statistical QoS constraints.

\item We investigate the interactions between the effective capacity, QoS constraints, channel sensing duration, channel detection threshold, false alarm and detection probabilities through numerical analysis.

\item We analyze both fixed-power/fixed-rate transmission schemes and variable schemes by considering different assumptions on the availability of channel side information (CSI) at the transmitter. We quantify the performance gains through power and rate adaptation.
\end{enumerate}


The organization of the rest of the paper is as follows. In Section \ref{sec:system}, we describe the
cognitive channel model. In Section \ref{sec:sensing}, we discuss channel sensing and provide expressions for the detection
and false alarm probabilities. In Section \ref{sec:transition1}, we describe the state transition model for cognitive transmission when the transmitter does not have CSI and sends the data at a fixed rate with fixed power, and we determine the effective capacity for this case. A similar analysis is conducted in Section \ref{sec:transition2} but now under the assumption that the transmitter has perfect CSI and employs power and rate adaptation. Finally, we provide conclusions in Section \ref{sec:conclusion}.

\section{System and Cognitive Channel Model} \label{sec:system}
We consider a cognitive radio channel model in which a secondary transmitter attempts to send information to a
secondary receiver possibly in the presence of primary users. Initially secondary users perform channel sensing,
and then depending on the primary users' activity, the secondary transmitter selects its transmission power and rate, i.e.,
when the channel is busy, the average power is $\overline{P}_{1}$ and the rate is $r_{1}$, and when the channel is
idle, the average power is $\overline{P}_{2}$ and the rate is $r_{2}$. For instance, if $\overline{P}_{1} = 0$, the secondary transmitter stops transmission in the presence of an active primary user. In the above model, the transmission rates $r_1$ and $r_2$ can be fixed or time-varying depending on whether the transmitter has channel side information or not. Moreover, in general we assume $\overline{P}_{1} < \overline{P}_{2}$.

We assume that the data generated by the source
is initially stored in the data buffer before being transmitted in frames of duration $T$ seconds over the cognitive
wireless channel. During transmission, the discrete-time channel input-output relation in the $i^{\text{th}}$ symbol duration is given by
\begin{align}\label{input-out1}
&y(i)=h(i)x(i)+n(i)\quad i=1,2,\dots
\end{align}
if the primary users are absent. On the other hand, if primary users are present in the channel, we have
\begin{align}\label{input-out2}
&y(i)=h(i)x(i)+s_{p}(i)+n(i)\quad i=1,2,\dots
\end{align}
Above, $x(i)$ and $y(i)$ denote the complex-valued channel input and output, respectively. We assume that the bandwidth
available in the system is $B$ and the channel input is subject to the following average energy constraints:
$\mathbb{E}\{|x(i)|^{2}\}\leq \overline{P}_{1}/B$ and $\mathbb{E}\{|x(i)|^{2}\}\leq \overline{P}_{2}/B$ for all $\textit{i}$,
when the channel is busy and idle, respectively. Since the bandwidth is $B$, symbol rate is assumed to be $B$
complex symbols per second, indicating that the average power of the system is constrained by $\overline{P}_{1}$ or $\overline{P}_{2}$.
In (\ref{input-out1}) and (\ref{input-out2}), $h(i)$ denotes the fading coefficient between the cognitive transmitter and the receiver. The fading coefficients can have arbitrary marginal distributions but they are assumed to have finite variances, i.e., $\mathbb{E}\{|h(i)|^2\} = \mathbb{E}\{z(i)\}= \sigma_h^2 < \infty$. Note that, here and throughout the paper, we have denoted the magnitude-square
of the fading coefficients by $z(i)=|h(i)|^{2}$. Finally, we consider a block-fading channel model and assume that the fading coefficients
stay constant for a block of duration $T$ seconds and change independently from one block to another independently.


In (\ref{input-out2}), $s_{p}(i)$ represents the sum of the active primary users' faded signals arriving at the secondary receiver.
In the input-output relations (\ref{input-out1}) and (\ref{input-out2}), $n(i)$ models the additive thermal noise at the receiver,
and is a zero-mean, circularly symmetric, complex Gaussian random variable with variance $\mathbb{E}\{|n(i)|^{2}\}=\sigma_{n}^{2}$
for all $i$. We further assume that $\{n_i\}$ is an independent and identically distributed (i.i.d.) sequence.

\section{Channel Sensing} \label{sec:sensing}
We assume that the first $N$ seconds of the frame duration $T$ is allocated to sense the channel. If the transmission strategies of the primary users are not known, energy-based detection methods are well-suited for the detection
of the activities of primary users.
The channel sensing can be formulated as a hypothesis testing problem between the noise $n(i)$ and the signal $s_p(i)$ in noise. Noting
that there are $NB$ complex symbols in a duration of $N$ seconds, this can mathematically be expressed as follows:
\begin{align}\label{hypothesis}
&\mathcal{H}_{0}\quad : \quad y(i)=n(i), \quad i=1,\dots,NB\\ \nonumber
&\mathcal{H}_{1}\quad : \quad y(i)=s_p(i)+n(i), \quad i=1,\dots,NB
\end{align}
Considering the above detection problem, the optimal Neyman-Pearson detector is given by \cite{Poor-book}
\begin{equation}\label{Neyman-Pearson}
Y=\frac{1}{NB}\sum_{i=1}^{NB}|y(i)|^{2}\gtrless^{\mathcal{H}_{1}}_{\mathcal{H}_{0}}\lambda
\end{equation}
where $\lambda$ is the detection threshold.
We assume that $s_p(i)$ has a circularly symmetric complex Gaussian distribution with zero-mean and variance $\sigma_{s_{p}}^{2}$.
Note that this is an accurate assumption if the signals are being received in a rich multipath environment or the number of active
primary users is large. Moreover, if, for instance the primary users are employing phase or frequency modulation, $s_p(i)$ in the
presence of even a single primary user in flat Rayleigh fading will be Gaussian distributed\footnote{Note that zero-mean, circular,
complex Gaussian distributions are invariant under rotation. For instance, if the fading coefficient $h$ is zero-mean, circularly symmetric, complex Gaussian distributed, then so is $he^{j\phi}$ for any random $\phi$.}. Assuming further that $\{s_p(i)\}$ are i.i.d., we can immediately
conclude that the test statistic $Y$ is chi-square distributed with $2NB$ degrees of freedom. In this case, the probabilities of false
alarm and detection can be established as follows:
\begin{align}\label{false alarm}
&P_{f}=Pr(Y>\lambda|\mathcal{H}_{0})=1-P\left(\frac{NB\lambda}{\sigma_{n}^{2}},NB\right)\\
&P_{d}=Pr(Y>\lambda|\mathcal{H}_{1})=1-P\left(\frac{NB\lambda}{\sigma_{n}^{2}+\sigma_{s_{p}}^{2}},NB\right) \label{eq:probdetect}
\end{align}
where $P(x,a)$ denotes the regularized lower gamma function and is defined as $P(x,a) = \frac{\gamma(x,a)}{\Gamma(a)}$ where $\gamma(x,a)$ is the lower incomplete gamma function and $\Gamma(a)$ is the Gamma function.

In the above hypothesis testing problem, another approach is to consider $Y$ as Gaussian distributed, which is accurate if $NB$ is large \cite{liang}. In this case, the detection and false alarm probabilities can be expressed in terms of Gaussian $Q$-functions. We would like to note that the rest of the analysis in the paper does not depend on the specific expressions of the false alarm and detection probabilities. However, numerical results are obtained using (\ref{false alarm}) and (\ref{eq:probdetect}).

\section{State Transition Model and Effective Capacity with CSI at the Receiver only} \label{sec:transition1}

In this section, we assume that the receiver has perfect channel side information (CSI)
and hence perfectly knows the instantaneous values of $\{h[i]\}$
while the transmitter has no such knowledge. Not knowing the channel conditions, the transmitter sends the information at fixed rates. More specifically, the transmission rate is fixed at $r_1$ bits/s in the presence of active primary users while the transmission rate is $r_2$ bits/s when the channel is idle. In this section, we initially construct a state-transition model for cognitive transmission by considering the cases in which the fixed transmission rates are smaller or greater than the instantaneous channel capacity values, and also incorporating the sensing decision and its correctness. In particular, if the fixed rate is smaller than the instantaneous channel capacity, we assume that reliable communication is achieved and the channel is in the ON state. Otherwise, we declare that outage has occurred and the channel is in the OFF state. Note that information has to be retransmitted in such a case. In the following, we provide a detailed description of the state transition model. Subsequently, we identify, through effective capacity, the maximum throughput that can be achieved in the described state-transition model when the system is subject to QoS constraints.

\subsection{State Transition Model} \label{subsec:state}

\begin{figure}
\begin{center}
\includegraphics[width = \ffigsize\textwidth]{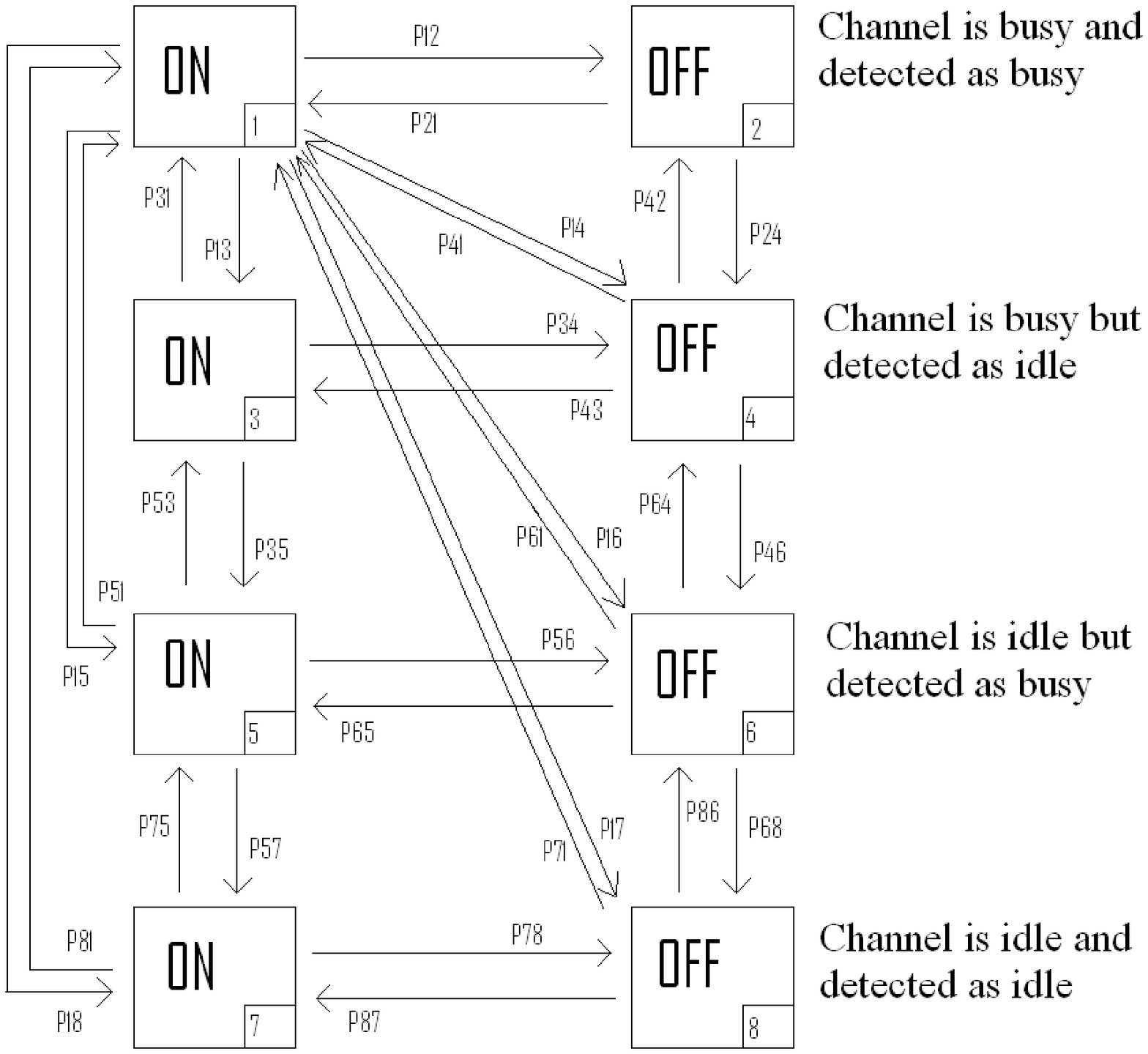}
\caption{State transition model for the cognitive radio channel. The numbered label for each state is given on the bottom-right corner of the box representing the state. } \label{fig:fig1}
\end{center}
\end{figure}

Regarding the decision of channel sensing and its correctness, we have the following four possible scenarios:
\begin{enumerate}
  \item Channel is busy, detected as busy (correct detection),
  \item Channel is busy, detected as idle (miss-detection),
  \item Channel is idle, detected as busy (false alarm),
  \item Channel is idle, detected as idle (correct detection).
\end{enumerate}
In each scenario, we have two states, namely ON and OFF, depending on whether or not the fixed-transmission rate exceeds the instantaneous channel capacity.
In order to identify these states, we have to first determine the instantaneous channel capacity values.
Note that if the channel is detected as busy, the secondary transmitter sends the information with power $\overline{P}_1$.
 Otherwise, it transmits with a larger power, $\overline{P}_2$. Considering
 the interference $s_p$ caused by the primary users
 as additional Gaussian noise, we can express the instantaneous channel capacities in the above four scenarios as follows:
\begin{align}\label{channel capacity}
&C_{1}=B\log_2(1+\tSNR_{1}z(i))\text{ (channel busy, detected busy)}\\
&C_{2}=B\log_2(1+\tSNR_{2}z(i))\text{ (channel busy, detected idle)}\\
&C_{3}=B\log_2(1+\tSNR_{3}z(i))\text{ (channel idle, detected busy)}\\
&C_{4}=B\log_2(1+\tSNR_{4}z(i))\text{ (channel idle, detected idle)}.
\end{align}
where $\tSNR_i$ for $i = 1,2,3,4$ denotes the average signal-to-noise ratio (SNR) values in each possible scenario. These SNR expressions are
\begin{equation} \label{eq:SNRs}
\tSNR_{1}=\frac{\overline{P}_{1}}{B\left(\sigma_{n}^{2}+\sigma_{s_{p}}^{2}\right)},\quad \textrm{} \quad
\tSNR_{2}=\frac{\overline{P}_{2}}{B\left(\sigma_{n}^{2}+\sigma_{s_{p}}^{2}\right)},\quad \textrm{} \quad
\tSNR_{3}=\frac{\overline{P}_{1}}{B\sigma_{n}^{2}},\quad \textrm{and} \quad \tSNR_{4}=\frac{\overline{P}_{2}}{B\sigma_{n}^{2}}.
\end{equation}
 Note that in scenarios 1 and 3, the channel is detected as busy and hence the transmission rate is $r_1$. On the other hand, the transmission rate is $r_2$ in scenarios 2 and 4. If these fixed rates are below the instantaneous capacity values, i.e., when $r_{1}< C_{1},C_{3}$ or $r_{2}< C_{2},C_{4}$, the cognitive transmission is considered to be in the ON state and reliable communication is achieved at these rates.
On the other hand, when $r_{1}\geq C_{1},C_{3}$ or $r_{2}\geq C_{2},C_{4}$, outage occurs and the transmission is in the OFF state. In this state, reliable communication is not attained, and hence, the information has to be resent. It is assumed that a simple automatic repeat request (ARQ) mechanism is incorporated in the communication protocol to acknowledge the reception of data and to ensure that erroneous data is retransmitted. This state-transition model with 8 states is depicted in Fig. \ref{fig:fig1} where the labels of the states are placed on the bottom-right corner. In states 1,3,5, and 7, the transmission is in the ON state, and $r_{1}(T-N)$ bits in states 1 and 5, and $r_{2}(T-N)$ bits in states 3 and 7 are transmitted and successfully received\footnote{Note that the transmission stays in each state for the frame duration of $T$ seconds. However, since $N$ seconds are allocated to channel sensing, data transmission occurs over a duration of $T-N$ seconds.}. The effective transmission rate is zero in the OFF states.

Next, we determine the state-transition probabilities. We use $p_{ij}$ to denote the transition probability from state $i$ to state $j$. Due to the block fading assumption, state transitions occur every $T$ seconds. When the channel is busy and detected as busy, the probability of staying in the ON state, which is topmost ON state in Fig. \ref{fig:fig1}, is expressed as follows:
\begin{align}
p_{11}&=\rho P_{d} \,P\{r_{1}<C_{1}(i+TB)\mid r_{1}<C_{1}(i)\}\\
&=\rho P_{d} \,P\{z(i+TB)>\alpha_{1}\mid z(i)>\alpha_{1}\} \label{eq:p11}
\end{align}
where
\begin{equation}
\alpha_{1}=\frac{2^{\frac{r_{1}}{B}}-1}{\tSNR_{1}},
\end{equation}
$\rho$ is the prior probability of channel being busy, and $P_{d}$ is the probability of detection as defined in (\ref{eq:probdetect}). Note that \eqref{eq:p11} is obtained under the assumption that the primary user activity is independent from frame to frame, leading to the expression which depends only on the prior probability $\rho$.
Note further that $p_{11}$ in general depends on the joint distribution of $(z(i+TB),z(i))$. However, since fading changes independently from one block to another in the block-fading model, we can further simplify  $p_{11}$ and write it as $$p_{11}=\rho P_{d}P\{z[i+TB]>\alpha_{1}\}=\rho P_{d}P\{z>\alpha_{1}\}$$ from which we can immediately see that the transition probability $p_{11}$ does not depend on the original state.  Hence, due to the block fading assumption, we can express
\begin{align}
p_{i1} = p_1 = \rho P_{d} P\{z > \alpha_1\} \quad \text{for } i = 1,2,\ldots,8. \label{eq:p1}
\end{align}
Similarly, the remaining transition probability expressions become
\begin{align}
&p_{i2} = p_{2}=\rho P_{d}P\{z < \alpha_1\}, \quad p_{i3} = p_{3}=\rho (1-P_{d})P\{z > \alpha_2\}, \\
&p_{i4} = p_{4}=\rho (1-P_{d})P\{z<\alpha_2\}, \quad p_{i5} = p_{5}=(1-\rho)P_{f}P\{z>\alpha_3\}, \\
&p_{i6} = p_{6}=(1-\rho)P_{f}P\{z<\alpha_3\}, \quad p_{i7} = p_{7}=(1-\rho)(1-P_{f})P\{z > \alpha_4\}, \\
&p_{i8} = p_{8}=(1-\rho)(1-P_{f})P\{z<\alpha_4\} \quad \text{for } i = 1,2,\ldots,8. \label{eq:p8}
\end{align}
where $\alpha_{2}=\frac{2^{\frac{r_{2}}{B}}-1}{\tSNR_{2}}$,
$\alpha_{3}=\frac{2^{\frac{r_{1}}{B}}-1}{\tSNR_{3}}$,
$\alpha_{4}=\frac{2^{\frac{r_{2}}{B}}-1}{\tSNR_{4}}$, and $P_f$ is the false alarm probability.
Now, the $8\times8$ state transition probability matrix can be expressed as
\begin{align}\label{eq:R}
R = \left[
    \begin{array}{ccccc}
        p_{1,1} & p_{1,2} & . & . & p_{1,8}\\
        . & . & .& . & .\\
        . & . &. & . & .\\
        p_{8,1} &p_{8,2} & . & . & p_{8,8}\\
    \end{array}
    \right]
     =\left[
    \begin{array}{ccccc}
        p_{1} &p_{2}& . & . & p_{8}\\
        . & . & .& . & .\\
        . &  .& .& . & .\\
        p_{1} & p_{2} & . & . & p_{8}\\
    \end{array}
\right].
\end{align}
Note that the rows of $R$ are identical, and therefore $R$ is a matrix of unit rank.

\subsection{Effective Capacity}
In this section, we identify the maximum throughput that the cognitive radio channel with the aforementioned state-transition model can sustain under statistical QoS constraints imposed in the form of buffer or delay violation probabilities. Wu and Negi in \cite{Wu} defined the effective capacity as the maximum constant arrival rate that can be supported by a given channel service process while also satisfying a statistical QoS requirement specified by the QoS exponent $\theta$. If we define $Q$ as the stationary queue length, then $\theta$ is defined as the decay rate of the tail distribution of the queue length $Q$:
\begin{equation}\label{decayrate}
\lim_{q\rightarrow \infty}\frac{\log P(Q\geq q)}{q}=-\theta.
\end{equation}
Hence, we have the following approximation for the buffer violation probability for large $q_{max}$: $P(Q\geq q_{max})\approx e^{-\theta q_{max}}$. Therefore, larger $\theta$ corresponds to more strict QoS constraints, while the smaller $\theta$ implies looser constraints. In certain settings, constraints on the queue length can be linked to limitations on the delay and hence delay-QoS constraints. It is shown in \cite{Liu} that $P\{D\geq d_{max}\}\leq c\sqrt{P\{Q\geq q_{max}\}}$ for constant arrival rates, where $D$ denotes the steady-state delay experienced in the buffer. In the above formulation, $c$ is a positive constant, $q_{max}=ad_{max}$ and $a$ is the source arrival rate. Therefore, effective capacity provides the maximum arrival rate when the system is subject to statistical queue length or delay constraints  in the forms of $P(Q \ge q_{\max}) \le
e^{-\theta q_{max}}$ or $P\{D \ge d_{\max}\} \le c \, e^{-\theta a \, d_{max}/2}$, respectively. Since the average
arrival rate is equal to the average departure rate when the queue
is in steady-state \cite{ChangZajic}, effective capacity can also be
seen as the maximum throughput in the presence of such constraints.

The effective capacity for a given QoS exponent $\theta$ is given by
\begin{equation}\label{exponent}
-\lim_{t\rightarrow \infty}\frac{1}{\theta t}\log_{e}\mathbb{E}\{e^{-\theta S(t)}\}=-\frac{\Lambda(-\theta)}{\theta}
\end{equation}
where $S(t)=\sum_{k=1}^{t}r(k)$ is the time-accumulated service process, and $\{r(k),k=1,2,\dots\}$ is defined as the discrete-time, stationary and ergodic stochastic service process. Note that the service rate is $r(k) = r_{1}(T-N)$ if the cognitive system is in state 1 or 5 at time $k$. Similarly, the service rate is $r(k) = r_{2}(T-N)$ in states 3 and 7. In all OFF states, fixed transmission rates exceed the instantaneous channel capacities and reliable communication is not possible. Therefore, the service rates in these states are effectively zero.

In the next result, we provide the effective capacity for the cognitive radio channel and state transition model described in the previous section.

\begin{theo} \label{theo:fixedrate}
For the cognitive radio channel with the state transition model given in Section \ref{subsec:state}, the normalized effective capacity in bits/s/Hz is given by
\begin{equation}
R_{E}(\tSNR,\theta)=\max_{r_{1},r_{2}\geq0} -\frac{1}{\theta TB} \log_{e} \bigg((p_{1}+p_{5})e^{-(T-N)\theta r_{1}}
+ (p_{3}+p_{7})e^{-(T-N)\theta r_{2}} + p_{2} + p_{4} + p_{6} + p_{8}\bigg) \label{eq:effectivecap}
\end{equation}
where $T$ is the frame duration over which the fading stays constant, $N$ is the sensing duration, $r_1$ and $r_2$ are fixed transmission rates, and $p_i$ for $i = 1, \ldots, 8$ are the transition probabilities expressed in \eqref{eq:p1}--\eqref{eq:p8}.
\end{theo}

\emph{Proof:} In \cite[Chap. 7, Example 7.2.7]{Performance}, it is shown for Markov modulated processes that
\begin{gather} \label{eq:theta-envelope}
\frac{\Lambda(\theta)}{\theta} = \frac{1}{\theta} \log_e sp(\phi(\theta)R)
\end{gather}
where $sp(\phi(\theta)R)$ is the spectral radius (i.e., the maximum of the absolute values of the eigenvalues) of the matrix $\phi(\theta)R$, $R$ is the transition matrix of the underlying Markov process, and $\phi(\theta) = \text{diag}(\phi_1(\theta), \ldots, \phi_M(\theta))$ is a diagonal matrix whose components are the moment generating functions of the processes in $M$ states. The rates supported by the cognitive radio channel with the state transition model described in the previous section can be seen as a Markov modulated process and hence the setup considered in \cite{Performance} can be immediately applied to our setting. Note that the transmission rates are non-random and fixed in each state in the cognitive channel. More specifically, the possible rates are $r_1(T-N)$, $r_2(T-N)$, and 0 for which the moment generating functions are $e^{\theta r_1(T-N)}$, $e^{\theta r_2(T-N)}$, and 1, respectively. Therefore, we have  $\phi(\theta) = \text{diag}\{e^{(T-N)\theta r_{1}},1,e^{(T-N)\theta r_{2}},1,e^{(T-N)\theta r_{1}},1,e^{(T-N)\theta r_{2}},1\}$. Then, using \eqref{eq:R}, we can write
\begin{align}
\phi(\theta)R=\left[
\begin{array}{ccccc}
\phi_{1}(\theta)p_{1} & .& . & . & \phi_{1}(\theta)p_{8}
\\
\phi_{2}(\theta)p_{1} & . & .& . & \phi_{2}(\theta)p_{8} \\
\phi_{3}(\theta)p_{1} & . & .& . & \phi_{3}(\theta)p_{8} \\
. &  . & .& .  & . \\
\phi_{8}(\theta)p_{1} & . & .& . & \phi_{8}(\theta)p_{8} \\
\end{array}
\right]
x=
\left[\begin{array}{ccccc}
e^{(T-N)\theta r_{1}} p_{1} & .& . & . & e^{(T-N)\theta r_{1}}p_{8}
\\
p_{1} & . & .& . & p_{1} \\
e^{(T-N)\theta r_{2}}p_{1} & . & .& . & e^{(T-N)\theta r_{2}}p_{8} \\
. &  . & .& .  & . \\
p_{8} & . & .& . & p_{8} \\
\end{array}
\right]
\end{align}
Since $\phi(\theta)R$ is a matrix with unit rank, we can readily find that
\begin{align}
sp(\phi(\theta)R) &= \text{trace}[\phi(\theta)R] = \phi_{1}(\theta)p_{1} +...+ \phi_{8}(\theta)p_{8}
\\
& = (p_{1} + p_5) e^{(T-N)\theta r_{1}}  + (p_{3}+p_7) e^{(T-N)\theta r_{2}} + p_2 + p_4 + p_{6} + p_{8}. \label{eq:sp}
\end{align}
Then, combining (\ref{eq:sp}) with (\ref{eq:theta-envelope}) and (\ref{exponent}), and noting that choice of the rates $r_1$ and $r_2$ can be optimized leads to the effective capacity formula given in (\ref{eq:effectivecap}). \hfill $\square$

We would like to note that the effective capacity expression in (\ref{eq:effectivecap}) is obtained for a given sensing duration $N$, detection threshold $\lambda$, and QoS exponent $\theta$. In the next section, we investigate the impact of these parameters on the effective capacity through numerical analysis.

\subsection{Numerical Results}

\begin{figure}
\begin{center}
\includegraphics[width = \figsize\textwidth]{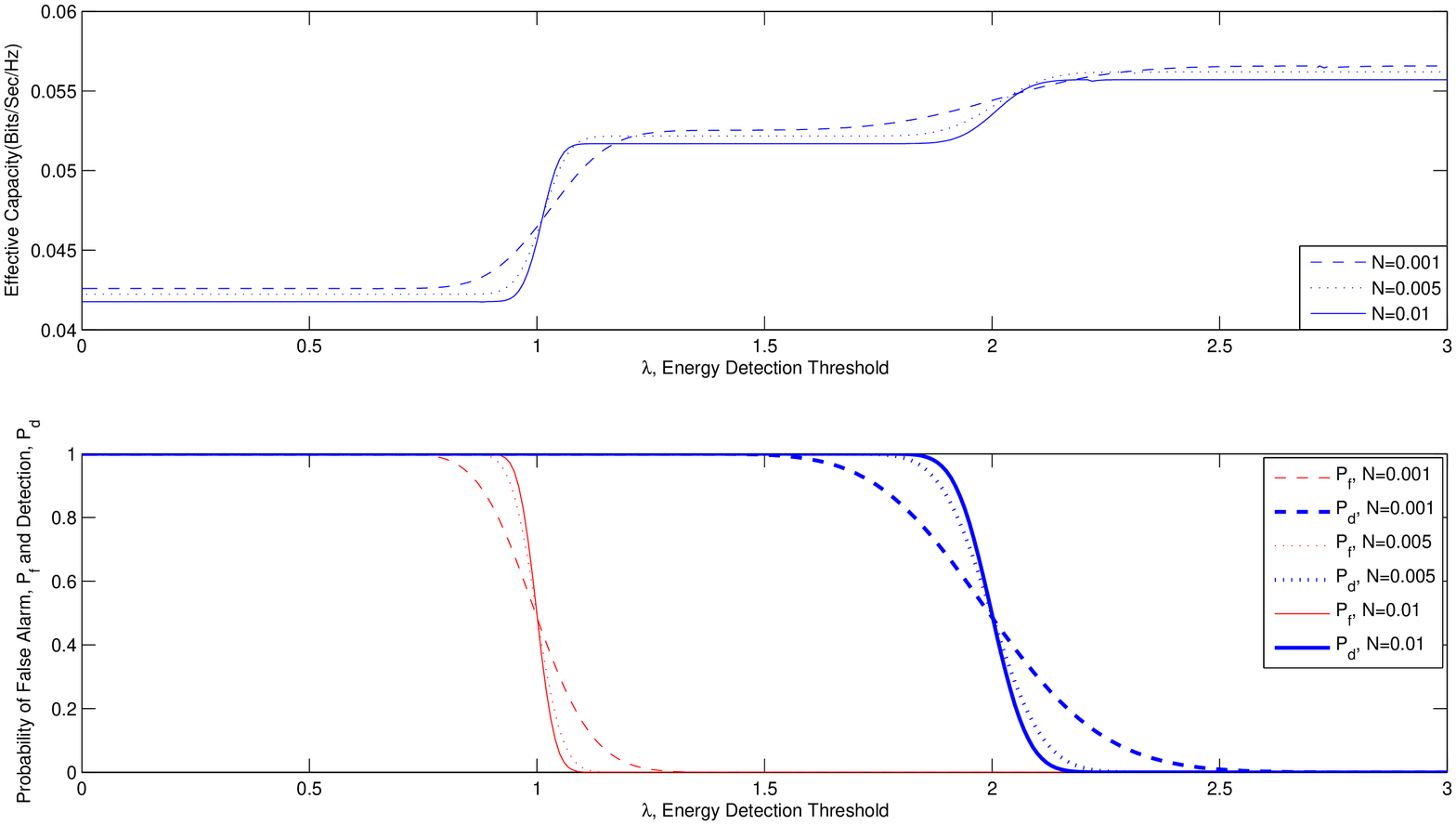}
\caption{Effective Capacity and $P_{f}-P_{d}$ v.s. Channel Detection Threshold $\lambda$. $\theta = 0.01$.} \label{fig:fig2}
\end{center}
\end{figure}

In this section, we present the numerical results. In Figure \ref{fig:fig2}, we plot the effective capacity as a function of the detection threshold value $\lambda$ for different sensing durations $N$. At the same time, we compare the false alarm and detection probabilities. The channel bandwidth is $100$kHz. We assume that the duration of the block is $T = 0.1$ seconds. The average input SNR values when the channel is detected correctly are $\tSNR_{1}=0$ dB and $\tSNR_{4}=10$ dB for busy and idle channels, respectively. The QoS exponent is $\theta=0.01$. The channel is assumed to be busy with an average probability of $\rho = 0.1$. As we see in Fig. \ref{fig:fig2},  the effective capacity is increasing with increasing $\lambda$. However, at the same time, as $\lambda$ increases, the probabilities of false alarm and detection are getting smaller. For instance, when $\lambda \approx 1$, the false alarm probabilities start diminishing, which in turn increases the effective capacity values significantly. If $\lambda$ is increased beyond 2, we observe that the detection probabilities start decreasing, causing increasing disturbance to the primary users. But, since the secondary user assumes that the channel is idle in the case of miss detection and transmits at a higher power level, we again see an increase in the effective capacity. Therefore, this increase occurs at the cost of increased interference to the primary users, which can be limited by imposing a lower bound on the detection probability. In Fig. \ref{fig:fig2}, we further observe that as the duration of channel sensing $N$ increases, the false alarm and detection probabilities decrease with sharper slopes. On the other hand, we note that having a larger $N$ decreases the effective capacity outside the range of $\lambda$ values at which transitions in the false alarm and detection probabilities occur. This is due to the fact that as $N$ increases, less time is available for data transmission.  Finally, we remark that if the threshold value $\lambda$ is taken between 1.2 and 1.7, the probabilities of false alarm and detection are 0 and 1, respectively, and the channel effective capacity is approximately 0.052 bits/sec/Hz. Such a favorable situation arises because of the large number of samples $NB$ used for channel sensing. If $B$ or $N$ is decreased significantly, false alarm and detection probabilities decrease with much smaller slopes, avoiding the possibility of realizing the above favorable scenario.

\begin{figure}
\begin{center}
\includegraphics[width = \figsize\textwidth]{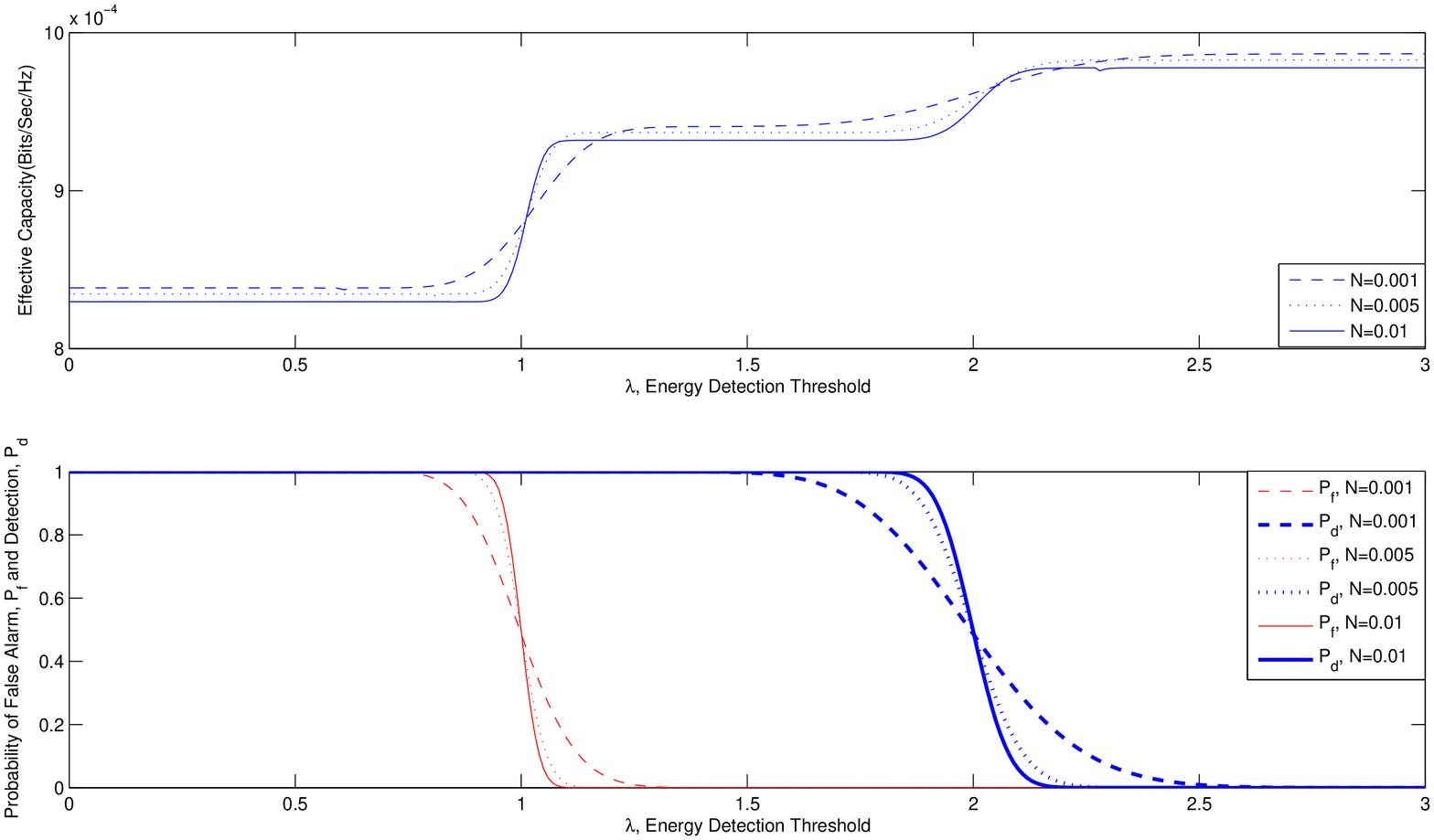}
\caption{Effective Capacity and $P_{f}-P_{d}$ v.s. Channel Detection Threshold $\lambda$. $\theta = 1$.} \label{fig:fig3}
\end{center}
\end{figure}

In Fig. \ref{fig:fig3}, all parameters other than $\theta$ are kept the same as the ones used in Fig. \ref{fig:fig2} while the QoS exponent is increased to $\theta=1$. Note that since the false alarm and detection probabilities do not depend on $\theta$, we have the same results as in Fig. \ref{fig:fig2}. Additionally, similar trends are observed in the effective capacity curves. However, since higher $\theta$ values mean more strict QoS limitations, we observe much smaller effective capacity values in Fig. \ref{fig:fig3}.



\begin{figure}
\begin{center}
\includegraphics[width = \figsize\textwidth]{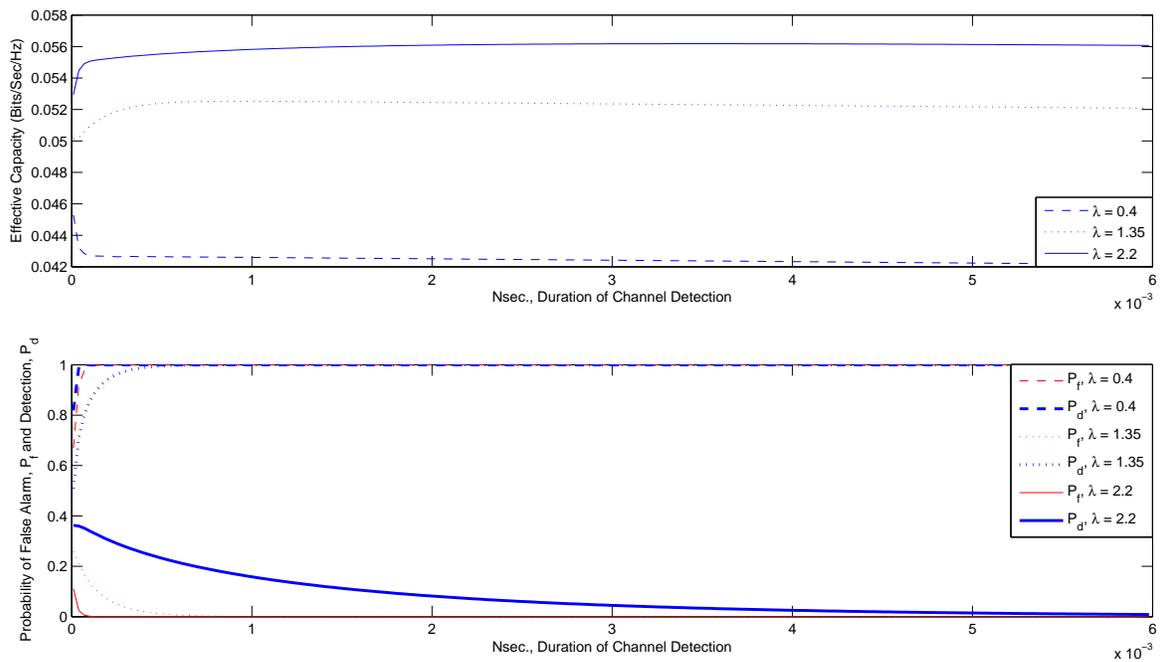}
\caption{Effective Capacity and $P_{f}-P_{d}$ v.s. Channel Sensing Duration, $N$. $\theta = 0.01$.} \label{fig:fig5}
\end{center}
\end{figure}

In Fig. \ref{fig:fig5}, we plot the effective capacity as a function of the  channel sensing duration, where we consider three different values of the channel detection threshold. The input SNR values are the same as the ones used in the previous figures and $\theta=0.01$. We observe that when $\lambda = 0.4$ and hence the threshold is small, both the false alarm and detection probabilities are high. Since false alarms happen frequently, effective capacity is small and gets smaller with increasing $N$. On the other hand, if $\lambda = 2.2$, false alarm and detection probabilities are low and decrease with increasing $N$. Hence, the secondary transmitter frequently assumes that the channel is idle and transmits with high power. As a result, the effective capacity is high. However, as remarked before, high interference is caused to the primary users. We further note that the effective capacity achieves its maximum value at $N \approx 0.0035$ above which it starts decreasing as less time is allocated to data transmission. When $\lambda = 1.35$, detection probabilities approach 1 and false alarm probabilities decrease to zero with increasing $N$. Hence, the channel is sensed reliably and disturbance to primary users is minimal. On the other hand, the effective capacity is smaller than that achieved when $\lambda = 2.2$.

\begin{figure}
\begin{center}
\includegraphics[width = \figsize\textwidth]{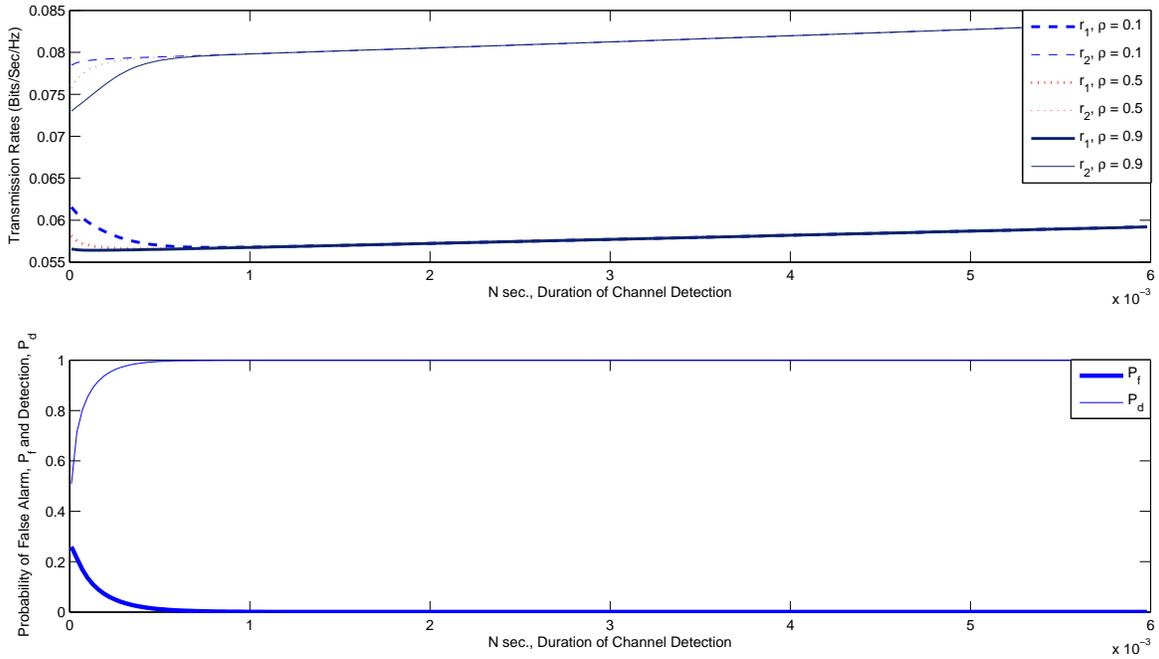}
\caption{Optimal Data Transmission Rates and $P_{f}-P_{d}$ v.s. Channel Sensing Duration $N$. $\theta = 0.01$.} \label{fig:fig6}
\end{center}
\end{figure}

In Fig. \ref{fig:fig6}, we plot the optimal transmission rates $r_1$ and $r_2$ with which the data is sent through the channel when the channel is busy and idle, respectively, as a function of the channel sensing duration $N$ for different values of channel occupancy probability $\rho$. We set $\lambda = 1.35$. As we can see, the optimal transmission rates $r_2$ for different values of $\rho$ converge when the detection probability is 1. Similarly, the optimal transmission rates $r_1$ for different values of the channel occupation probabilities converge when the false alarm probability is 0. Hence, the optimal rates are independent of $\rho$ when the false alarm and detection probabilities are 0 and 1, respectively.

\begin{figure}
\begin{center}
\includegraphics[width = \figsize\textwidth]{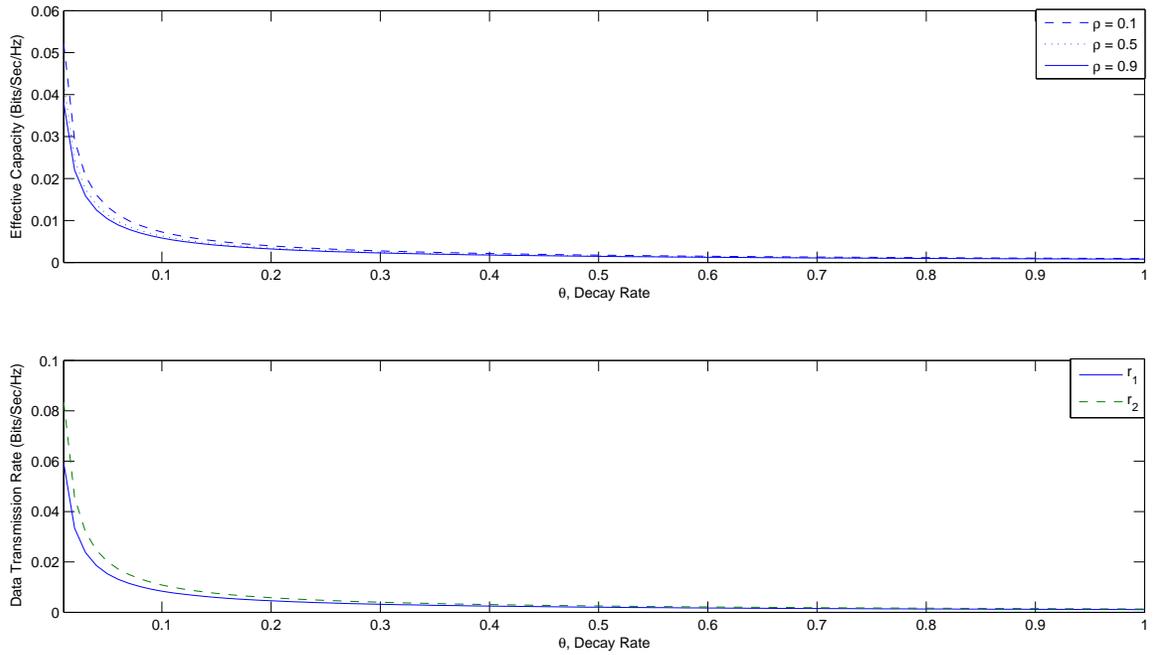}
\caption{Effective Capacity and Optimal Data Transmission Rates v.s. QoS exponent $\theta$} \label{fig:fig7}
\end{center}
\end{figure}

In Fig. \ref{fig:fig7}, with the assumption that the primary users' activities are known perfectly (i.e., there is no sensing error), we display the effective capacity and optimal data transmission rates obtained at different channel occupation probabilities $\rho$ as a function of the QoS exponent $\theta$. In the upper part of the figure, we notice that the effective capacity is decreasing with increasing $\theta$ and increasing primary user activity in the channel. We also observe that as $\theta$ increases and hence more strict QoS are imposed, the sensitivity of the effective capacity to $\rho$ decreases. In the lower part of Fig. \ref{fig:fig7}, we plot the optimal data transmission rates. The dashed line shows the rates when the channel is empty whereas the solid line gives the rates used when the channel is occupied by the primary users. Here, we observe that while the optimal data transmission rates are decreasing with increasing $\theta$, they are independent of $\rho$ and hence the primary users' activity in the channel.

\section{State Transition Model and Effective Capacity with CSI at Both the Receiver and Transmitter} \label{sec:transition2}

In this section, we assume that both the transmitter and the receiver have perfect CSI, and hence perfectly know the instantaneous values of $\{h[i]\}$. With this assumption, as a major difference from Section \ref{sec:transition1}, we now allow the transmitter to adapt its rate and power with respect to the channel conditions. In particular, we assume that the transmitter sends the information at the rate that is equal to the instantaneous channel capacity value, and employs the normalized power adaptation policies $\mu_{1}(\theta,z(i)) = \frac{P_1(\theta,z(i))}{\overline{P}_1}$ when the channel is busy, and $\mu_{2}(\theta,z(i)) = \frac{P_2(\theta,z(i))}{\overline{P}_2}$ when the channel is idle. Note that the power adaptation schemes are normalized by the average power constraints $\overline{P}_1$ and $\overline{P}_2$, and they depend on the QoS exponent $\theta$ and the instantaneous channel state $z(i) = |h(i)|^2$. Note further that the power adaptation policies need to satisfy the average power constraints:
\begin{equation}\label{mu}
\mathbb{E}_z\{\mu_{1}(\theta,z))\} = \int_{0}^{\infty}\mu_{1}(\theta,z))f(z)\textit{d}z\leq 1 \quad \textrm{and} \quad \mathbb{E}_z\{\mu_{2}(\theta,z))\} = \int_{0}^{\infty}\mu_{2}(\theta,z))f(z)\textit{d}z\leq 1
\end{equation}
where $f(z)$ denotes the probability density function (pdf) of $z = |h|^2$.

\subsection{State Transition Model} \label{subsec:state1}
 With respect to the decision of channel sensing, we still have the four possible channel scenarios outlined at the beginning of Section \ref{subsec:state}. Below, we specify the instantaneous capacity values and the corresponding rates used by the transmitter in each possible scenario:
 \begin{align}
  &\hspace{-0.5cm}C_{1}(i)=B\log_2(1+\mu_{1}(\theta,z(i))z(i)\tSNR_{1}), \quad r_{1}(i)=C_{1}(i); \quad \text{Channel busy, detected as busy.} \quad \text{The channel is ON}. \nonumber
  \\
  &\hspace{-0.5cm}C_{2}(i)=B\log_2(1+\mu_{2}(\theta,z(i))z(i)\tSNR_{2}), \quad r_{2}(i)>C_{2}(i); \quad \text{Channel busy, detected as idle.} \quad \text{The channel is OFF}. \nonumber
  \\
  &\hspace{-0.5cm}C_{3}(i)=B\log_2(1+\mu_{1}(\theta,z(i))z(i)\tSNR_{3}), \quad r_{1}(i)<C_{3}(i); \quad \text{Channel idle, detected as busy.} \quad  \text{The channel is ON}. \nonumber
  \\
  &\hspace{-0.5cm}C_{4}(i)=B\log_2(1+\mu_{2}(\theta,z(i))z(i)\tSNR_{4}), \quad r_{2}(i)=C_{4}(i); \quad \text{Channel idle, detected as idle.} \quad \text{The channel is ON}. \nonumber
\end{align}
where SNR expressions are the same as those defined in \eqref{eq:SNRs}.
Note that, in contrast to the analysis in Section \ref{subsec:state}, we in this section have only one state (either ON or OFF) for each scenario. If the channel is detected as busy, the secondary transmitter sends the data at the instantaneous rate
\begin{equation}\label{rate1}
r_{1}(i)=B\log_2(1+\mu_{1}(\theta,z(i))z(i)\tSNR_{1})
\end{equation}
where $\mu_{1}(\theta,z(i))$ is the power adaptation policy in this case. Depending on the channel's true state being busy or idle (scenarios 1 or 3 above), $r_1(i)$ is either equal to the instantaneous channel capacity as in scenario 1 or less than that as in scenario 3. Hence, in both cases, reliable transmission can be attained at the rate of $r_1(i)$, and the channels are ON. When the channel is detected as idle, the data transmission rate is
\begin{equation}\label{rate2}
r_{2}(i)=B\log_2(1+\mu_{2}(\theta,z(i))z(i)\tSNR_{4}).
\end{equation}
If the channel is actually idle, $r_2(i)$ is equal to the instantaneous channel capacity, and therefore the channel is in the ON state as in scenario 4. On the other hand, if the channel is busy but detected as idle as in scenario 2 above, $r_2(i)$ is greater than the channel capacity because the transmitter does not take into account the interference caused by the primary users. Hence, this becomes the only case in which the channel is in the OFF state. Similarly as before, we assume that outage occurs in this state and reliable transmission cannot be provided. The information has to be resent with the assistance of an ARQ mechanism.

In summary, we have three ON states and one OFF state under the assumptions of this section. These states correspond to states 1, 4, 5, and 7 of Fig. \ref{fig:fig1}. Therefore, the state transition model in this section can be obtained by keeping these states and eliminating states 2, 3, 6, and 8 in the state-transition model in Fig. \ref{fig:fig1}. Note that as another major difference from the state-transition model in Section \ref{subsec:state}, the transmission rates in each state are now random processes. Therefore, in this new model, the state-transition probabilities depend only on the detection probabilities and the prior probability of channel being busy, $\rho$. These probabilities can be expressed as
\begin{align}
p_{i1} = p_1 = \rho P_d, \quad p_{i4} = p_4 = \rho(1-P_d), \quad p_{i5} = p_5 = (1-\rho)P_f, \text{ and } p_{i7} = p_7 = (1-\rho)(1-P_f),
\end{align}
for $i = 1,4,5,$ and 7.


\subsection{Effective Capacity}
The following result provides the effective capacity expression when the transmitter, having perfect CSI, employs rate and power adaptation during transmission.
\begin{theo}\label{theo:variableratepower}
For the cognitive radio channel with power and rate adaptation at the transmitter and with the state transition model described in Section \ref{subsec:state1}, the normalized effective capacity in bits/s/Hz is given by
\vspace{-.3cm}
\begin{align}
R_E(\tSNR,\theta)=\max_{\substack{\mu_{1}(\theta,z):\E_z\{\mu_{1}(\theta,z)\}\le 1 \\ \mu_{2}(\theta,z):\E_z\{\mu_{2}(\theta,z)\}\le 1}} -\frac{1}{\theta TB}\log_e\bigg[&\left(\rho P_{d}
+(1-\rho)P_{f}\right)\E_z\{e^{-(T-N)\theta r_{1}}\} \nonumber
\\
&+(1-\rho)(1-P_{f})\E_z\{e^{-(T-N)\theta r_{2}}\}+\rho(1-P_{d})\bigg] \label{maximized}
\end{align}
where the expectations are with respect to $z$, and $r_1 = B\log_2(1+\mu_{1}(\theta,z)z\tSNR_{1})$ and $r_2 = B\log_2(1+\mu_{2}(\theta,z)z\tSNR_{4})$.
\end{theo}

\emph{Proof}: The proof is very similar to the proof of Theorem \ref{theo:fixedrate}. The only differences are that we now have four states and the service processes (or equivalently the transmission rates) are random processes that depend on $z$. As described in Section \ref{subsec:state1}, the rates are $r_1(i)$ in states 1 and 5, $r_2(i)$ in state 7, and zero in state 4. Therefore, the corresponding moment generating functions are $\phi_1(\theta) = \phi_5(\theta) = \mathbb{E}_z\{e^{(T-N)\theta r_1}\}$, $\phi_7(\theta) = \mathbb{E}_z\{e^{(T-N)\theta r_2}\}$, and $\phi_4(\theta) = 1$, where the expectations are with respect to $z$. Using the same approach as in the proof of Theorem \ref{theo:fixedrate}, we can easily find that
\begin{align}
\frac{\Lambda(\theta)}{\theta} &= \frac{1}{\theta} \log_e \left( (p_{1} + p_5) \mathbb{E}_z\{e^{(T-N)\theta r_1}\}  + p_7 \mathbb{E}_z\{e^{(T-N)\theta r_2}\} + p_4 \right)
\\
&= \frac{1}{\theta} \log_e \left( (\rho P_d + (1-\rho)P_f) \mathbb{E}_z\{e^{(T-N)\theta r_1}\}  + (1-\rho)(1-P_f) \mathbb{E}_z\{e^{(T-N)\theta r_2}\} + \rho(1-P_d) \right). \label{eq:lambda}
\end{align}
Combining the expression in \eqref{eq:lambda} with \eqref{exponent}, and maximizing over all possible power adaptation schemes leads to \eqref{maximized}. \hfill $\square$

Having obtained the expression for the effective capacity, we now derive the optimal power adaptation strategies that maximize the effective capacity.
\begin{theo} \label{theo:optimalpower}
The optimal power adaptation policies that maximize the effective capacity are given by
\begin{equation}\label{mu1}
\mu_{1}(\theta,z)=\left\{
          \begin{array}{ll}
            \frac{1}{\tSNR_{1}}\left(\frac{1}{\gamma_1^{\frac{1}{a+1}}}\frac{1}{z^{\frac{a}{a+1}}}-\frac{1}{z}\right), & \hbox{$z>\gamma_1$} \\
            0, & \hbox{otherwise}
          \end{array}
        \right.
\end{equation}
and
\begin{equation}\label{mu2}
\mu_{2}(\theta,z)=\left\{
          \begin{array}{ll}
            \frac{1}{\tSNR_{4}}\left(\frac{1}{\gamma_{2}^{\frac{1}{a+1}}}\frac{1}{z^{\frac{a}{a+1}}}-\frac{1}{z}\right), & \hbox{$z>\gamma_{2}$} \\
            0, & \hbox{otherwise.}
          \end{array}
        \right.
\end{equation}
where $a=(T-N)B\theta/\log_e2$. $\gamma_{1}$ and $\gamma_{2}$ are the threshold values in the power adaptation policies and they can be found from the average power constraints in (\ref{mu}) through numerical techniques.
\end{theo}
\emph{Proof:} Since logarithm is a monotonic function,  the optimal power adaptation policies can also be obtained from the following minimization problem
\begin{align}\label{objective}
\min_{\substack{\mu_{1}(\theta,z):\E_z\{\mu_{1}(\theta,z)\}\le 1 \\ \mu_{2}(\theta,z):\E_z\{\mu_{2}(\theta,z)\}\le 1}} \left(\rho P_{d}
+(1-\rho)P_{f}\right)\E_z\{e^{-(T-N)\theta r_{1}}\}+(1-\rho)(1-P_{f})\E_z\{e^{-(T-N)\theta r_{2}}\}.
\end{align}
It is clear that the objective function in (\ref{objective}) is strictly convex and the constraint functions in (\ref{mu}) are linear with respect to $\mu_{1}(\theta,z)$ and $\mu_{2}(\theta,z)$. Then, forming the Lagrangian function and setting the derivatives of the Lagrangian with respect to $\mu_{1}(\theta,z)$ and $\mu_{2}(\theta,z)$ equal to zero, we obtain
\begin{align}\label{lagran1}
&\left\{\lambda_{1}-a\tSNR_{1}z\left[\rho P_{d}+(1-\rho)P_{f}\right]\left[1+\mu_{1}(\theta,z)z\tSNR_{1}\right]^{-a-1}\right\}f(z)=0\\
&\left\{\lambda_{2}-a\tSNR_{4}z\left(1-\rho\right)\left(1-P_{f}\right)\left[1+\mu_{2}(\theta,z)z\tSNR_{4}\right]^{-a-1}\right\}f(z)=0 \label{lagran2}
\end{align}
where $\lambda_1$ and $\lambda_2$ are the Lagrange multipliers.
Defining $\gamma_{1}=\frac{\lambda_{1}}{\left[\rho P_{d}+(1-\rho)P_{f}\right]a\tSNR_{1}}$ and
$\gamma_{2}=\frac{\lambda_{2}}{\left(1-\rho\right)\left(1-P_{f}\right)a\tSNR_{4}}$,
and solving (\ref{lagran1}) and \eqref{lagran2}, we obtain optimal power policies
given in (\ref{mu1}) and (\ref{mu2}). 
\hfill $\square$

The optimal power allocation schemes identified in Theorem \ref{theo:optimalpower} are similar to that given in \cite{zhang}. However, in the cognitive radio channel, we have two allocation schemes depending on the presence or absence of active primary users. Note that the optimal power allocation in the presence of active users, $\mu_{1}(\theta,z(i)) = \frac{P_1(\theta,z(i))}{\overline{P}_1}$, has to be performed under a more strict average power constraint since $\overline{P}_1 < \overline{P}_2$. Note also that under certain fading conditions, we might have $\mu_{1}(\theta,z(i)) > \overline{P}_1$, causing more interference to the primary users. Therefore, it is also of interest to apply only rate adaptation and use fixed-power transmission in which case we have $\mu_{1}(\theta,z(i)) = \mu_{2}(\theta,z(i)) = 1$. We can immediately see from the result of Theorem \ref{theo:variableratepower} that the effective capacity of fixed-power/variable-rate transmission is
\begin{align}
R_E(\tSNR,\theta)= -\frac{1}{\theta TB}\log_e\bigg[&\left(\rho P_{d}
+(1-\rho)P_{f}\right)\E_z\{e^{-(T-N)\theta r_{1}}\} +(1-\rho)(1-P_{f})\E_z\{e^{-(T-N)\theta r_{2}}\}+\rho(1-P_{d})\bigg]
\end{align}
where $r_1 = B\log_2(1+z\tSNR_{1})$ and $r_2 = B\log_2(1+z\tSNR_{4})$.


\subsection{Numerical Results}

\begin{figure}
\begin{center}
\includegraphics[width = \figsize\textwidth]{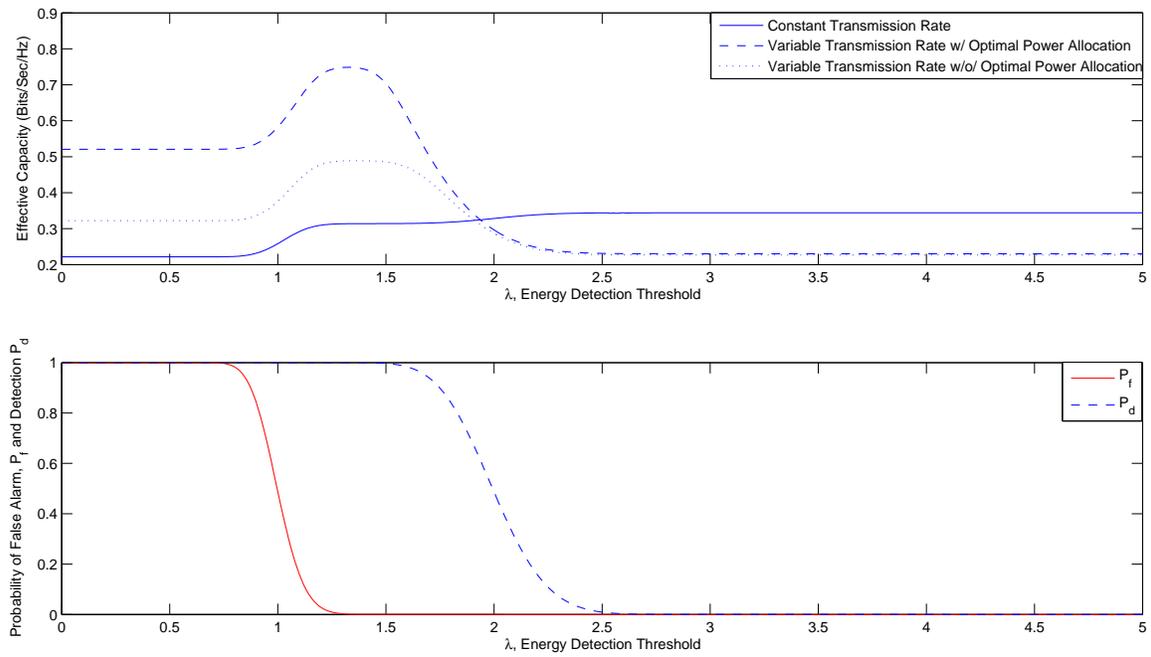}
\caption{Effective Capacity and $P_{f}-P_{d}$ for different schemes v.s. Energy Detection Threshold, $\lambda$. $\theta = 0.01$.} \label{fig:fig8}
\end{center}
\end{figure}

In Fig. \ref{fig:fig8}, we plot the effective capacities of the three transmission schemes, namely, fixed-power/fixed-rate transmission (solid line), variable-power/variable-rate transmission (dashed-line), and fixed-power/variable-rate transmission (dotted-line), discussed heretofore in the paper, as a function of the detection threshold $\lambda$. We note that the optimal power adaptation is employed in the variable-power scheme. In this figure, all the parameters are the same as in Fig. \ref{fig:fig2} discussed in Section \ref{sec:transition1}. Hence, $\theta = 0.01$. When we compare variable-rate/variable-power and variable-rate/fixed-power schemes, we immediately notice, as expected, that variable-rate/variable-power outperforms the latter one for all $\lambda$ values. However, the difference in the effective capacity values reduces as $\lambda$ is increased beyond $\approx$2 where detection probability starts diminishing. Additionally, we observe that for $\lambda  \ge 2$, fixed-rate/fixed-power scheme starts outperforming the variable schemes. Note that when the detection probability is small, miss-detections occur frequently. In variable schemes, recall that the transmission enters the OFF state in cases of miss-detection in which the channel is detected as idle but is actually busy, and hence a degradation in the performance is expected. This is also the reason for why the effective capacity of the variable schemes is decreasing for $\lambda$ values greater than 1.5 where the detection probability has also started getting smaller than 1. Note that this is in stark contrast to the behavior exhibited by the fixed-rate/fixed-power scheme. We finally note that the variable schemes perform better than the fixed-rate/fixed-power transmission when the detection probabilities are relatively high (or equivalently when $\lambda < \approx 2$), and also as before, an decrease in the false alarm probability increases the rates.

\begin{figure}
\begin{center}
\includegraphics[width = 0.8\textwidth]{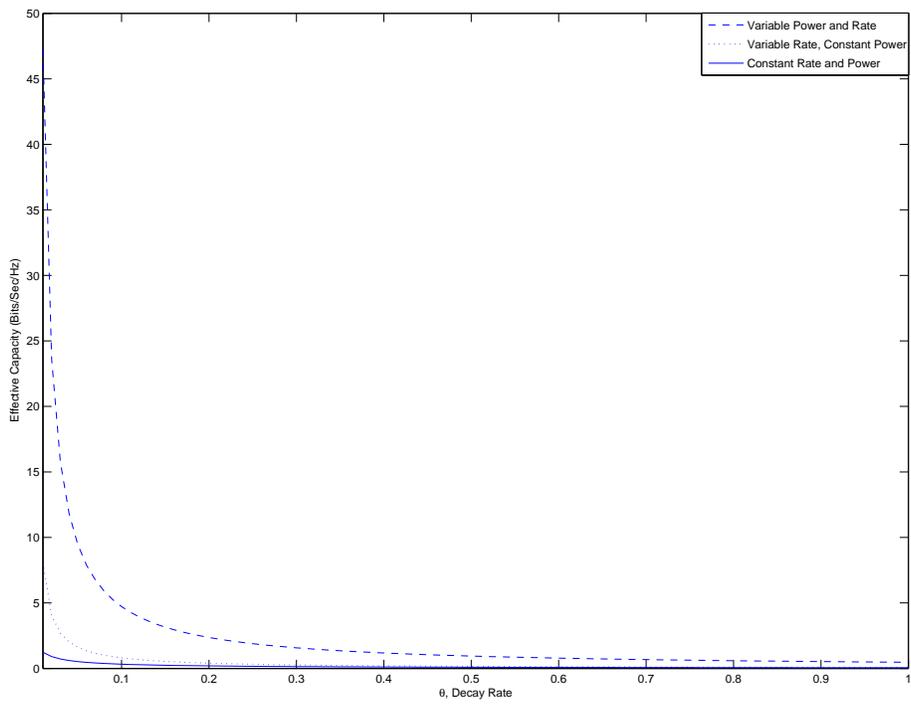}
\caption{Effective Capacity for different schemes v.s. QoS exponent $\theta$} \label{fig:fig9}
\end{center}
\end{figure}

Fig. \ref{fig:fig9} plots the effective capacities of different transmission schemes as a function of the QoS exponent $\theta$ under the assumption of perfect channel detection i.e., the probability of false alarm is 0 and the probability of detection is 1. As expected, effective capacity values are decreasing with increasing $\theta$ values. Since the plot is obtained under perfect channel sensing, the transmission strategy with variable power and rate outperforms the other two schemes for all $\theta$ values. However, we interestingly note that the gains attained through adapting the power and rate tend to diminish with increasing $\theta$. Hence, QoS constraints have a significant impact in this respect.

\vspace{-.2cm}
\section{Conclusion} \label{sec:conclusion}
In this paper, we have analyzed the effective capacity
of cognitive radio channels in order to identify the performance levels and to determine the interactions between throughput and channel sensing parameters in the presence of QoS constraints. We have initially constructed a state-transition model for cognitive transmission and then obtained expressions for the effective capacity. This analysis is conducted for fixed-power/fixed-rate, fixed-power/variable-rate, and variable-power/variable-rate transmission schemes under different assumptions on the availability of CSI at the transmitter. Through numerical results, we have investigated the impact of channel sensing duration and threshold, detection and false alarm probabilities, and QoS limitations on the throughput. Several insightful observations are made. We have noted that the effective capacity in general increases with decreasing false alarm probabilities. On the other hand, we have remarked that having smaller detection probabilities have a different effect in fixed-rate and variable-rate schemes. We have seen that variable schemes outperform fixed-rate transmission methods if the detection probabilities are sufficiently high.  Otherwise, fixed-power/fixed-rate transmission should be preferred.  We have observed that both the effective capacity and transmission rates get smaller with increasing $\theta$. We have also noted that the gains through adapting rate and power diminish as $\theta$ increases and hence as QoS constraints become more stringent.
\end{spacing}
\vspace{-0.1cm}
\begin{spacing}{1.35}

\end{spacing}

\newpage

\end{document}